\documentclass[aps,pra,superscriptaddress,amsmath,amssymb,preprintnumbers,showpacs,twocolumn]{revtex4}

\usepackage{amssymb}
\usepackage{graphicx}
\usepackage{bm}
\usepackage{empheq}
\usepackage{color}
\usepackage{upgreek}
\usepackage{cancel}
\usepackage{slashed}

\newcommand{\Ignore}[1]{}

\newcommand{\Ket}[1]{\left\vert #1\right\rangle}

\def\e{\mathrm{e}}
\def\ii{\mathrm{i}}
\def\d{\mathrm{d}}

\begin{document}

\title{Interaction-free evolution in the presence of time-dependent Hamiltonians}

\author{Dariusz Chru\'sci\'nski}
\address{Institute of Physics, Faculty of Physics, Astronomy and Informatics Nicolaus Copernicus University, Grudziadzka 5/7, 87–100 Torun, Poland}
\author{Antonino Messina}
\address{Dipartimento di Fisica e Chimica, Universit\`a degli Studi di Palermo, Via Archirafi 36, I-90123 Palermo, Italy}
\author{Benedetto Militello}
\address{Dipartimento di Fisica e Chimica, Universit\`a degli Studi di Palermo, Via Archirafi 36, I-90123 Palermo, Italy}
\author{Anna Napoli}
\address{Dipartimento di Fisica e Chimica, Universit\`a degli Studi di Palermo, Via Archirafi 36, I-90123 Palermo, Italy}

\begin{abstract}

The generalization of the concept of interaction-free evolutions
(IFE) [A. Napoli, {\it et al.}, Phys. Rev. A {\bf 89}, 062104
(2014)] to the case of time-dependent Hamiltonians is discussed.
It turns out that the time-dependent case allows for much more
rich structures of interaction-free states and interaction-free
subspaces.  The general condition for the occurrence of IFE is
found and exploited to analyze specific situations. Several
examples are presented, each one associated to a class of
Hamiltonians with specific features.

\end{abstract}


\maketitle

\section{Introduction}\label{sec:Introduction}

An interaction-free evolution (IFE) of a quantum system is an
evolution which is not influenced by a certain part of the
Hamiltonian which is addressed as the interaction
term~\cite{ref:IFE}. In other words, the dynamics generated by the
\lq unperturbed\rq\, Hamiltonian $H_0$ is essentially the same as
the evolution generated by the total Hamiltonian which is the sum
of $H_0$ and the interaction term $H_{\rm I}$: $H=H_0 + H_{\rm
I}$. This notion, which has been introduced in
Ref.~\cite{ref:IFE}, is somehow related to the concept of
decoherence-free subspaces
(DFS)~\cite{ref:IFE_PS,ref:zanardi1,ref:zanardi2,ref:Lidar,ref:Lidar2013}.
In spite of such connection, it should be stressed  that the two
concepts are still different in many aspects. Generally speaking,
the notion of IFE can be relevant to composite systems with
different dimensions (like a small system and its environment) or
with similar dimensions (for example two interacting qubits), but
it can even concern different degrees of freedom of the same
particle (for example atomic and vibrational degrees of freedom of
a trapped ion). One can even talk about IFE states in connection
with the action of a classical field on a quantum system, for
example a spin under the action of a magnetic field.

Subradiance~\cite{ref:subradiant1,ref:subradiant2,ref:subradiant3,ref:subradiant4,ref:subradiant5,ref:BMN,ref:BM},
in its original formulation, is surely a very famous phenomenon
which can be thought of as an IFE involving a matter system
(several atoms) and the vacuum electromagnetic field.

In this paper, we study the non trivial extension of IFE states
which applies to those cases wherein the system is governed by a
time-dependent Hamiltonian. The interest in such a kind of problem
is related to several aspects. On the one hand, generally speaking
the resolution of dynamical problems with time-dependent
Hamiltonians is a tough job due to the highly nontrivial structure
of the corresponding solution
\begin{equation}\label{Ut}
  U(t) = {\cal T} \exp\left(-\ii \int_0^t H(\tau)\d\tau \right) ,
\end{equation}
where $\mathcal{T}$ denotes the chronological product. In general
Eq.~(\ref{Ut}) is untractable and, except for some lucky
cases~\cite{ref:TDH1,ref:TDH2,ref:TDH3}, it requires special
assumptions, such as for example the adiabatic
one~\cite{ref:TDH4}, or suitable approximations, like in the
perturbative treatment~\cite{ref:TDH5,ref:TDH6}. Therefore, even
the partial resolution of a class of time-dependent problems in
the presence of time-dependent Hamiltonians is of interest itself.
Formula (\ref{Ut}) simplifies if $H(t)$ defines a commutative
family, i.e. $[H(t),H(t')]=0$ for arbitrary $t$ and $t'$. In this
case the chronological product drops out and the entire evolution
is controlled by the integral $\int_0^t H(\tau)\d\tau$.

On the other hand, there could be important applications in the
field of quantum control and in particular in the field of
suppression of decoherence effects. Indeed, our analysis, could
pave the way to extensions of the concepts of subradiance and
decoherence-free subspaces in the presence of time-dependent
Hamiltonian of the system and even in the presence of
time-dependent interaction between the system and its environment.

The paper is organized as follows. In the next section we
introduce the problem and find out the general conditions that
guarantee the interaction-free evolution. In sections
\ref{sec:FirstExamples}, \ref{sec:Adiabatic} and
\ref{sec:ProperTD} we provide several examples of IFE states
belonging to different classes. In particular, after the simplest
examples in sec.~\ref{sec:FirstExamples}, we go on, in
sec.~\ref{sec:Adiabatic}, by analyzing a case of IFE in the
context of an adiabatic evolution, while in
sec.~\ref{sec:ProperTD} we present some examples related to a more
general class of IFE states. Finally, in
sec.~\ref{sec:Discussion}, we give some conclusive remarks.

\section{Interaction-Free Conditions}\label{sec:general}

Let us recall the definition of interaction-free evolution (IFE):
we say that a state $\Ket{\psi_\mathrm{0}}$ undergoes an IFE if it
evolves as if the interaction term of the Hamiltonian (which can
be time-dependent) were absent. To better understand this
definition, let us assume that our system is governed by a
time-dependent Hamiltonian which can be split into two parts: one
part that we call unperturbed and one part that we call
interaction term:
\begin{equation}\label{eq:Schrodinger}
  \ii\partial_t \Ket{\psi(t)} =
  (H_\mathrm{0}(t)+H_\mathrm{I}(t))\Ket{\psi(t)}\,.
\end{equation}
The relevant evolution operator is denoted by $U(t)$, while
$U_\mathrm{0}(t)$ denotes the evolution operator associated to
$H_\mathrm{0}(t)$ only. Which means,
\begin{subequations}
\begin{eqnarray}
  \label{eq:UTot}
  \ii\partial_t U(t) &=& (H_\mathrm{0}(t)+H_\mathrm{I}(t))U(t)\,,\\
  \label{eq:UZero}
  \ii\partial_t U_\mathrm{0}(t) &=&
  H_\mathrm{0}(t)U_\mathrm{0}(t)\,.
\end{eqnarray}
\end{subequations}
We are looking for those states $\Ket{\psi_\mathrm{0}} \in
\mathcal{H}$ (the Hilert space of the system) for which the complete evolution is \lq
essentially\rq\,  equal to the unperturbed one:
\begin{equation}\label{eq:IFE-Def}
 U(t)\Ket{\psi_\mathrm{0}} = \e^{\ii
  A(t)}U_\mathrm{0}(t)\Ket{\psi_\mathrm{0}}\,,
\end{equation}
where $A(t)$ is a real function of time.

Inserting  the ansatz $\Ket{\psi(t)}=\e^{\ii
A(t)}U_\mathrm{0}(t)\Ket{\psi_\mathrm{0}}$ into the Schr\"odinger
equation in Eq.~(\ref{eq:Schrodinger}), and exploiting
Eqs.~(\ref{eq:UTot}) and (\ref{eq:UZero}), we get that the
following condition must be satisfied:
\begin{equation}\label{eq:MainCondition}
  (H_\mathrm{I}(t) - \dot{A}(t))U_\mathrm{0}(t)\Ket{\psi_\mathrm{0}} = 0\,,
\end{equation}
which means that, at every instant, the unperturbed evolution
operator $U_0(t)$  maps the initial state $\Ket{\psi_\mathrm{0}}$
into an instantaneous eigenstate of the interaction term:
\begin{equation}\label{HU=aU}
  H_\mathrm{I}(t)U_0(t)\Ket{\psi_\mathrm{0}} = a(t)\,U_\mathrm{0}(t)\Ket{\psi_\mathrm{0}} \,,
\end{equation}
with $a(t) = \dot{A}(t)$. This condition is clearly necessary and
sufficient, since the chain of implications that brings from
Eq.~(\ref{eq:IFE-Def}) to Eq.~(\ref{HU=aU}) can be followed
backward, from Eq.~(\ref{HU=aU}) to Eq.~(\ref{eq:IFE-Def}).

It is worth mentioning that all the states satisfying
Eq.~(\ref{HU=aU}) with the same $a(t)$ form a subspace, that we
will address as an IFE subspace. In fact, every state belonging to
such a subspace evolves as if the interaction were not present. On
the contrary, if one considers the superposition of two IFE states
belonging to different IFE subspaces, a phase difference between
such states will be accumulated (due to the different values of
the eigenvalue $a(t)$), and then the evolution will be effectively
different from the one obtained in the absence of interaction.

Let us observe that applying $U^\dag_\mathrm{0}(t)$ to the both
sides of Eq.~(\ref{HU=aU})  one gets:
\begin{equation}  \label{INT}
  \tilde{H}_\mathrm{I}(t)\Ket{\psi_\mathrm{0}} = a(t)\Ket{\psi_\mathrm{0}} \,,
\end{equation}
where $\tilde{H}_\mathrm{I}(t) = U_0^\dagger(t) H_\mathrm{I}(t)
U_0(t)$ is the interaction term in the interaction picture. This
means that the initial state $\Ket{\psi_\mathrm{0}}$ is supposed
to be an eigenstate of $\tilde{H}_\mathrm{I}(t)$ for all $t$. It
should be stressed that $\Ket{\psi_\mathrm{0}}$ being an
eigenvector of $\tilde{H}_\mathrm{I}(t)$ does not need to be an
eigenvector of $H_\mathrm{I}(t)$, which is clear from
Eq.~(\ref{HU=aU}). Note, however, that if  $\Ket{\psi_\mathrm{0}}$
satisfies
\begin{equation}\label{S0}
  H_\mathrm{I}(t)\Ket{\psi_\mathrm{0}} = a(t)\,\Ket{\psi_\mathrm{0}} \,,
\end{equation}
and
\begin{equation}\label{S1}
  [H_\mathrm{I}(t) - a(t) \mathbb{I}] H_\mathrm{0}(t_1)H_\mathrm{0}(t_2) \ldots H_\mathrm{0}(t_n) \Ket{\psi_\mathrm{0}} = 0 \,,
\end{equation}
for $n=1,2,\ldots$, then (\ref{HU=aU}) is surely satisfied (cf.
Appendix A).  It should be stressed that Eqs.~(\ref{S0}) and
(\ref{S1}) are only sufficient but not necessary conditions for
$\Ket{\psi_\mathrm{0}}$ to be an IFE state. The condition in
Eq.~(\ref{INT}) (as well as that in Eq.~(\ref{HU=aU})) is both
necessary and sufficient for $\Ket{\psi_\mathrm{0}}$ to be IFE
state.

Interestingly, in the time-independent case they reduce to
\begin{equation}\label{S}
  [H_\mathrm{I} -  a \,\mathbb{I}] H_\mathrm{0}^n \Ket{\psi_\mathrm{0}} = 0 \,,
\end{equation}
for $n=1,2,\ldots,N-1$, where $N = {\rm dim}\, \mathcal{H}$. It
was proved \cite{ref:IFE} that these conditions are both
necessary and sufficient. It is, therefore, clear that time-dependent case is
much more complicated and rich showing that
$\Ket{\psi_\mathrm{0}}$ needs not to be eigenvector of
$H_\mathrm{I}(t)$ for $t \neq 0$, but
$U_0(t)\Ket{\psi_\mathrm{0}}$ must belong to an eigenspace of the
interaction Hamiltonian $H_\mathrm{I}(t)$ at any time.

On the basis of Eq.~(\ref{INT}) we can distinguish between two
possible situations where the interaction picture interaction term
is time-dependent or not. Nevertheless, in order to be effective,
such a classification should explore in detail also a sort of \lq
grey zone\rq\, which corresponds to all those cases where the
Hamiltonian has a trivial time dependence, like for example
$\tilde{H}_\mathrm{I}(t)=f(t)\tilde{H}_\mathrm{I}(0)$ (we will
provide several examples of this kind). Though we will not go
through such a {\it taxonomic} approach, in the examples given in
the following sections we will always comment on the specific
relevant properties of $\tilde{H}_\mathrm{I}(t)$.

\section{Single Systems subjected to external fields}\label{sec:FirstExamples}

As a class of time-dependent Hamiltonians that allow then
occurrence of interaction-free evolutions we will consider the
cases of magnetic moments immersed in suitable magnetic fields.

\subsection{Spin-$1/2$ particle}

Let us consider a spin--$1/2$ particle immersed in a time-dependent magnetic
field. The corresponding Hamiltonian is expressible as follows:
\begin{equation}
  H(t) = - \mu \mathbf{B}(t)\cdot \mathbf{S}\,,
\end{equation}
where $\mathbf{S} =(\sigma_x,\sigma_y,\sigma_z)$.
Then we take the $z$ contribution, for the moment assumed
time-independent, as the unperturbed Hamiltonian, and the rest as
the interaction term ($\hbar=1$):
\begin{subequations}
\begin{equation}
  H_\mathrm{0}(t) = \frac{\Omega}{2}\sigma_z\,,
\end{equation}
\begin{eqnarray}
  H_\mathrm{I}(t) &=& \alpha(t)[\cos(\Omega t + \phi)\sigma_x + \sin(\Omega t + \phi)\sigma_y]\,.
\end{eqnarray}
\end{subequations}
We introduce the notation
$\sigma_\theta=\cos\theta\sigma_x+\sin\theta\sigma_y$. The
corresponding eigenvectors of  $\sigma_\theta$ read:
\begin{equation}
  \Ket{\pm}_\theta = \frac{1}{\sqrt{2}}\left(\e^{-\ii\theta/2}\Ket{+} \pm  \e^{\ii\theta/2}\Ket{-}\right) \,,
\end{equation}
where $\Ket{\pm}$ are the eigenstates of $\sigma_z$.

Now, suppose that the initial state $\Ket{\psi_\mathrm{0}}$ is an
eigenstate of the operator $\sigma_\phi$: $\Ket{\psi_\mathrm{0}} =
\Ket{\pm}_\phi$. It is easy to show that the evolution operator
associated to the unperturbed Hamiltonian, which is nothing but a
rotation along the $z$ axis, maps such an initial state into an
instantaneous eigenstate of $H_\mathrm{I}(t)$:
\begin{eqnarray}
 &&  U_0(t) \Ket{\pm}_\phi = \Ket{\pm}_{\Omega t + \phi} = \nonumber   \\
&&  \frac{1}{\sqrt{2}}\left(\e^{-\ii(\Omega t + \phi)/2}\Ket{+} \pm \e^{\ii(\Omega t + \phi)/2}\Ket{-}\right) .
\end{eqnarray}
In such a case the total evolution is essentially given by the
unperturbed evolution, up to a phase factor:
\begin{eqnarray}
  \Ket{\pm}_\phi &\rightarrow& U(t)\Ket{\pm}_\phi = \e^{\mp \ii A(t)} U_\mathrm{0}(t)\Ket{\pm}_{\phi}\nonumber \\
   &=& \e^{\mp \ii A(t)} \Ket{\pm}_{\Omega t + \phi} \,,
\end{eqnarray}
with $A(t) = \int_0^t \alpha(s)ds$. Of course in each subspace a
different phase due to $H_\mathrm{I}$ is accumulated.

It is worth noting that we are beyond the trivial case where
$H_\mathrm{0}$ and $H_\mathrm{I}$ commute. In fact, they don't
commute at all, but the operator $U_\mathrm{0}(t)$ maps
eigenstates of $H_\mathrm{I}(0)$ into eigenstates of
$H_\mathrm{I}(t)$.

This results are still valid if we generalize the Hamiltonian
model:
\begin{eqnarray}
  H(t) &=& H_0(t) + H_{\rm I}(t) \\
  &=&  \frac{\Omega(t)}{2}\sigma_z +
  \alpha(t)[ \cos(\Phi(t))\sigma_x+\alpha(t)\sin(\Phi(t))\sigma_y] , \nonumber
\end{eqnarray}
with
\begin{equation}\label{}
 \Phi(t) = \int_0^t \Omega(s)\mathrm{d}s + \phi \,.
\end{equation}

There is a clear physical interpretation in terms of classical
counterpart of such behaviours. We have a magnetic moment
$\mathbf{m}$ on the $xy$ plane which is rotating under the action
of a magnetic field along $z$. Now we add another magnetic field
of the $xy$ plane, say $\mathbf{B}_\perp(t)$ which is always
parallel to the magnetic moment. At any instant of time, the
component $\mathbf{B}_\perp$ does not act on the spin, since the
relevant torque is vanishing ($\tau =
\mathbf{m}\times\mathbf{B}_\perp = 0$), and then the presence of
$\mathbf{B}_\perp$ does not affect the motion of the spin.

It should be clear that if $H_\mathrm{I} = \alpha
[\cos\phi\sigma_x + \sin\phi\sigma_y]$ does not depend on time,
then there is no interaction-free state corresponding to
$H_\mathrm{0} = \frac 12 \Omega(t) \sigma_z$. This shows in a
clear way the difference between time-independent and time
dependent cases.

\subsection{Spin-$1$ particle}

Let us now consider a {\it toy model} involving spin-$1$ operators
(cf. Appendix B). After introducing the following notation,
\begin{equation}
  L_\phi = \cos\phi L_x + \sin\phi L_y\,,
\end{equation}
we consider the following Hamiltonian:
\begin{subequations}
\begin{equation}
  H(t) = \Omega(t) L_z + \alpha(t) L_{\phi(t)}^2\,,
\end{equation}
with
\begin{equation}
  \phi(t) = \int_0^t \Omega(s)\mathrm{d}s + \phi(0) \,.
\end{equation}
\end{subequations}
As the initial condition we take the state
\begin{equation}
  \Ket{\psi_\mathrm{0}}=c_-\Ket{-1}_{\phi(0)}+c_+\Ket{+1}_{\phi(0)}\,,
\end{equation}
with
\begin{equation}
  \Ket{\pm 1}_{\phi} =
  \frac{\e^{-\ii\phi}}{2}\Ket{+1}\pm\frac{1}{\sqrt{2}}\Ket{0}+\frac{\e^{\ii\phi}}{2}\Ket{-1}
  \,,
\end{equation}
and $\Ket{-1}$, $\Ket{0}$, $\Ket{+1}$ the eigenstates of $L_z$ in
the subspace with $l=1$.

This is an example where the unperturbed Hamiltonian maps an
eigenspace of the interaction Hamiltonian at the initial time to
the corresponding eigenspace of the interaction Hamiltonian at
time $t$.  In fact, the operator $L_{\phi(t)}^2$ has a twofold
degenerate subspace corresponding to the eigenvalue $1$ and a
singlet corresponding to zero. This means that the two states
$\Ket{-1}_{\phi(t)}$ and $\Ket{+1}_{\phi(t)}$ do not \lq feel\rq\,
the interaction Hamiltonian except for the (same) phase
accumulated, which is $\e^{-\ii \int_0^t\alpha(s)\mathrm{d}s}$.

It deserves to be noted that the examples in this section are such
that the relevant interaction Hamiltonian in the interaction
picture provides a commutative family of operators, i.e, it has
the following form $\tilde{H}_\mathrm{I}(t) = f(t)
\tilde{H}_\mathrm{I}(0)$. In fact, for spin-1/2 we have:
\begin{equation}\label{}
  \tilde{H}_\mathrm{I}(t) = \alpha(t) \left( \begin{array}{cc}  0 & e^{-i \phi} \\ e^{i\phi} & 0 \end{array} \right)\ ,
\end{equation}
and hence it has time-independent eigenvectors $\Ket{\pm}_\phi$
and time-dependent eigenvalues $\pm \alpha(t)$. For spin-1 one
finds:
\begin{equation}\label{}
   \tilde{H}_\mathrm{I}(t) = \alpha(t) \left( \begin{array}{ccc} 1 &  0 & e^{-2i \phi(0)}  \\ 0 & 2 & 0 \\ e^{2i \phi(0)}  &  0 & 1 \end{array} \right) \,
\end{equation}
which has a \lq static\rq\, doublet corresponding to the
eigenvalue $1$.

\section{Adiabatic evolutions}\label{sec:Adiabatic}

Also adiabatic evolutions can provide interesting examples of
interaction-free evolutions, though approximated. Consider the
Hamiltonian of the class used for Stimulated Raman Adiabatic
Passage
(STIRAP)~\cite{ref:STIRAP1,ref:STIRAP2,ref:STIRAP3,ref:STIRAP4}.
The unperturbed Hamiltonian in the basis $\Ket{1}$, $\Ket{2}$,
$\Ket{3}$ reads:
\begin{eqnarray}
  H_\mathrm{0}(t) = \left(
    \begin{array}{ccc}
      0 & \Omega\sin\theta(t) & 0 \\
      \Omega\sin\theta(t) & \Delta & \Omega\cos\theta(t) \\
      0 & \Omega\cos\theta(t) & 0
    \end{array}
  \right) .
\end{eqnarray}
The three instantaneous eigenvalues of $H_\mathrm{0}$ are given by
\begin{eqnarray}
  \lambda = 0\, , \,\, \frac{\Delta \pm \sqrt{\Delta^2+4\Omega^2}}{2} \,.
\end{eqnarray}
The instantaneous eigenstate corresponding to the zero eigenvalue reads
\begin{eqnarray}
  \Ket{v(t)} = \cos\theta(t) \Ket{1} - \sin\theta(t) \Ket{3} \,.
\end{eqnarray}
In the adiabatic limit, assuming $\theta(0)=0$ and
$\theta(\infty)=\pi/2$, one has that the state $\Ket{1}$ is
adiabatically mapped into $\Ket{3}$. This is the essence of the
counterintuitive STIRAP sequence.

Consider now the following additional interaction term:
\begin{eqnarray}
  \nonumber
  H_\mathrm{I}(t) = \epsilon(t)\left(
    \begin{array}{ccc}
      \cos^2\theta(t) & 0 & -\sin\theta(t)\cos\theta(t) \\
      0 & 0 & 0 \\
      -\sin\theta(t)\cos\theta(t) & 0 & \sin^2\theta(t)
    \end{array}
  \right)\,.\\
\end{eqnarray}

It consists of a direct interaction between the states $\Ket{1}$
and $\Ket{3}$ and two shifts of the levels involved in such an
interaction.

The state $\Ket{v(t)}$ is an instantaneous eigenstate of the
interaction term, corresponding to the eigenvalue $\epsilon(t)$.
Therefore, in the adiabatic limit associated to the change of
$H_\mathrm{0}(t)$, the state $\Ket{v(0)}$ is mapped into
$\Ket{v(t)}$, which does not feel $H_\mathrm{I}(t)$, except for
the accumulation of a dynamical phase.

Of course, in this case the result is only approximated, since the
adiabatic evolution is only an approximation of the complete
evolution induced by $H_\mathrm{0}(t)$.

Similarly to the examples given in the previous section, even in
this example that we have provided for adiabatic evolutions the
eigenstates of $\tilde{H}_\mathrm{I}(t)$ do not change. Indeed,
since $v|(t)\>$ is common instantaneous eigenstate of
$H_\mathrm{0}(t)$ and $H_\mathrm{I}(t)$, then it turns out that
$|v(0)\>$ is eigenstate of $\tilde{H}_\mathrm{I}(t)$ at every
time, corresponding to the eigenvalue $\epsilon(t)$, and the
remaining subspace is the kernel of $\tilde{H}_\mathrm{I}(t)$, and
then $\tilde{H}_\mathrm{I}(t) = \epsilon(t)/\epsilon(0)
\tilde{H}_\mathrm{I}(0)$.

\section{Essential Time-Dependence of $\tilde{H}_\mathrm{I}$}\label{sec:ProperTD}

Since all the examples given in the previous sections are
related to those cases where $\tilde{H}_\mathrm{I}(t)$ has a
trivial time-dependence, in this section we provide some examples
of \textit{real} time-dependent $\tilde{H}_\mathrm{I}(t)$ which
have some time-independent eigenstates.

\subsection{The multi-photon nonlinear JC model}

The following Hamiltonian,
\begin{equation}
  H(t) = \omega \hat{n} + \frac{\Omega}{2}\sigma_z + \gamma \left[ e^{-\ii (\Omega - k\omega - \Delta)t} f(\hat{n}) \hat{a}^k \sigma_+ + h.c. \right]\,,
\end{equation}
can be obtained for example in the physical scenario of trapped
ions subjected to a laser slightly off-resonant to the $k$-th red
sideband ($\omega_L = \Omega - k\omega - \Delta$), out of the
Lamb-Dicke limit (which implies the presence of the \lq
coefficient\rq \, $f(\hat{n})$) and in the
RWA\cite{ref:trappedions}.

Taking,
\begin{subequations}
\begin{eqnarray}
  H_\mathrm{0} &=& \omega \hat{n} + \frac{\Omega}{2}\sigma_z\,,\\
  H_\mathrm{I}(t) &=& \gamma \left[ e^{\ii (\Omega - k\omega - \Delta)t} f(\hat{n}) \hat{a}^k \sigma_+ + h.c. \right]\,,
\end{eqnarray}
\end{subequations}
one can easily prove that,
\begin{eqnarray}
  \tilde{H}_\mathrm{I}(t) = \gamma \left[ e^{- \ii \Delta t} f(\hat{n}) \hat{a}^k \sigma_+ + h.c. \right]\,,
\end{eqnarray}
and that the multiplet $\{\Ket{0,g}$, $\Ket{1,g}$, \ldots,
$\Ket{k-1,g}\}$ (with $\sigma_z\Ket{g}=-\Ket{g}$) defines an
eigenspace of $\tilde{H}_\mathrm{I}(t)$. Of course, it is not an
eigenspace of $H_\mathrm{0}$, which implies that, though it is
interaction-free, in this subspace there could be a non trivial
evolution due to the action of $H_\mathrm{0}$.

Note that the interaction term in the interaction picture
$\tilde{H}_\mathrm{I}(t)$ in this case is time-dependent, though
it has a time-independent eigenspace (its kernel).

\subsection{Sum of multi-photon JC models}

Also the following Hamiltonian can be obtained in trapped ions
scenario:
\begin{equation}
  H(t) = \omega \hat{n} + \frac{\Omega}{2}\sigma_z + \left[ (\gamma_k(t) \hat{a}^k + \gamma_l(t) \hat{a}^l ) \sigma_+ + h.c.
  \right]\,.
\end{equation}
The time-dependence of the coupling parameters $\gamma$'s can be
realized through a modulation of the amplitudes of the laser
fields.

Let us assume that $k>l$. If $\gamma_l=0$ then the kernel of the interaction
Hamiltonian is generated by all the states $\Ket{m, g}$ with $m=0,
1, ..., k-1$, while in the other case we have only the states with
$m=0, 1, ..., l-1$. Therefore, in the case where $\gamma_l(t)$
changes and vanishes at some instants of time, the kernel of
$H_\mathrm{I}$ changes, but some states always belong to it. Such
states ($m\le l-1$) and all their linear combinations undergo
interaction-free evolution.

These two examples can be properly generalized considering for
example $\Omega(t)$ instead of a time-independent $\Omega$, in
order to have a time-dependent $H_\mathrm{0}$.

\section{Discussion}\label{sec:Discussion}

In this paper we have generalized the concept of IFE to the case
of time-dependent Hamiltonians. We have first of all provided
necessary and sufficient conditions for such an occurrence. Then,
we have presented several examples, related to different possible
structures of the system under scrutiny. The very first examples
(spin-$1/2$ and spin-$1$) analyze small quantum systems
interacting with time-dependent classical fields. In particular,
in the case of spin-$1$ we discuss the case where an IFE
eigenspace is present (the doublet corresponding to angular
momentum projections equal to $-1$ and $+1$). In the subsequent
example we have considered IFE states in the presence of an
adiabatic evolution, especially in the context of STIRAP. Finally,
in section \ref{sec:ProperTD} we have considered two cases of
spin-boson interaction (for example the vibrational and electronic
degrees of freedom of a trapped ion). In such a situation, we have
two interacting subsystems each one not feeling the interaction
with the other, if the total system is prepared in suitable (IFE)
states. Moreover, in one case, the IFE subspace has dimension
varying in time.

On the basis of the analysis developed in section
\ref{sec:general}, we know that the more compact condition to find
out IFE subspaces is that IFE states are nothing but states which
are  eigenstates of the interaction-picture interaction
Hamiltonian at every time instant, which really clarify the
physical origin of the dynamical features of such states.

At this point, it is worth to mention that the concept of IFE
states (whether with time-independent or time-dependent
Hamiltonian), when applied to a system interacting with its
environment, has some connection with the concept of
decoherence-free subspaces, as already pointed out in Ref.
\cite{ref:IFE}. Nevertheless, reporting on a detailed analysis of
the relation between IFE and DFS is beyond the scope of this paper
and will be presented elsewhere.

\section*{Acknowledgements}

We thank the anonymous referee for valuable comments. DC was
partially supported by the National Science Center project
DEC-2011/03/B/ST2/00136.

\section*{Appendix A}

The evolution operator $U_\mathrm{0}(t)$ associated to the
unperturbed Hamiltonian $H_\mathrm{0}(t)$ can be expanded as:
\begin{eqnarray}
 U_\mathrm{0}(t)&&=\mathbb{I}-i \int_0^t H_\mathrm{0}(t_1)dt_1+\\ \nonumber
 &&+ (-i)^2\int_0^t dt_1 \int_0^{t_1} H_\mathrm{0}(t_1)H_\mathrm{0}(t_2)dt_2+...
\end{eqnarray}
Thus,
\begin{eqnarray}\label{A1}
 &&H_\mathrm{I}(t)U_\mathrm{0}(t)\Ket{\psi_\mathrm{0}}=\\ \nonumber && H_\mathrm{I}(t)\Ket{\psi_\mathrm{0}}-i \int_0^t H_\mathrm{I}(t)H_\mathrm{0}(t_1)dt_1\Ket{\psi_\mathrm{0}}+\\
 \nonumber
 &&+ (-i)^2\int_0^t dt_1 \int_0^{t_1}
H_\mathrm{I}(t)H_\mathrm{0}(t_1)H_\mathrm{0}(t_2)dt_2\Ket{\psi_\mathrm{0}}+...
\end{eqnarray}
Starting from Eq.~(\ref{A1}) it is immediate to convince oneself
that if  $\Ket{\psi_\mathrm{0}} $ satisfies
Eqs.~(\ref{S0})--(\ref{S1}) then
\begin{equation}
  H_\mathrm{I}(t)U_0(t)\Ket{\psi_\mathrm{0}} = a(t)\,U_\mathrm{0}(t)\Ket{\psi_\mathrm{0}} \,,
\end{equation}
that is $\Ket{\psi_\mathrm{0}}$ is an IFE state.

\section*{Appendix B}

The spin-1 operators are defined as follows:
\begin{subequations}
\begin{equation}
  L_x = \frac{1}{\sqrt{2}} \left( \begin{array}{ccc} 0 & 1 & 0 \\ 1 & 0 & 1 \\   0 & 1 & 0 \end{array} \right) \ , \ \
\end{equation}
\begin{equation}
  L_y = \frac{1}{\sqrt{2}} \left( \begin{array}{ccc} 0 & -i & 0 \\ i & 0 & -i \\   0 & i & 0 \end{array} \right) \ ,
\end{equation}
\begin{equation}
L_z = \left( \begin{array}{ccc} 1 & 0 & 0 \\ 0 & 0 & 0 \\   0 & 0
& -1 \end{array} \right)\ .
\end{equation}
\end{subequations}

The operator $L_\phi$ has eigenvalues $\{0,1,-1\}$ corresponding
to the following eigenstates:
\begin{subequations}
\begin{equation}
  \Ket{0}_{\phi} =
  \frac{\e^{-\ii\phi}}{2}\Ket{+1}-\frac{\e^{\ii\phi}}{2}\Ket{-1}
  \,,
\end{equation}
\begin{equation}
  \Ket{\pm 1}_{\phi} =
  \frac{\e^{-\ii\phi}}{2}\Ket{+1}\pm\frac{1}{\sqrt{2}}\Ket{0}+\frac{\e^{\ii\phi}}{2}\Ket{-1}
  \,.
\end{equation}
\end{subequations}

Its square,
\begin{equation}
L^2_\phi = \frac 12 \left( \begin{array}{ccc} 1 & 0 & e^{-2i\phi}
\\ 0 & 2 & 0 \\   e^{2i\phi} & 0 & 1 \end{array} \right)\,,
\end{equation}
has the same eigenstates and the following eigenvalues: $0$
(singlet) and $1$ (doublet).

\end{document}